\documentclass[sigconf]{acmart}

\AtBeginDocument{
  \providecommand\BibTeX{{
    \normalfont B\kern-0.5em{\scshape i\kern-0.25em b}\kern-0.8em\TeX}}}

\setcopyright{rightsretained}
\copyrightyear{2023}
\acmYear{2023}
\acmConference[UIST '23]{The 36th Annual ACM Symposium on User Interface Software and Technology}{October 29-November 1, 2023}{San Francisco, CA, USA}
\acmBooktitle{The 36th Annual ACM Symposium on User Interface Software and Technology (UIST '23), October 29-November 1, 2023, San Francisco, CA, USA}
\acmDOI{10.1145/3586183.3606751}
\acmISBN{979-8-4007-0132-0/23/10}

\begin{document}
%!TEX root = main.tex
\definecolor{darkgreen}{rgb}{0,0.5,0}
\definecolor{orange}{rgb}{1,0.5,0}
\definecolor{teal}{rgb}{0,0.5,0.5}
\definecolor{darkpurple}{rgb}{0.5, 0, 0.5}

%use these commands while writing
% \newcommand {\bjoern}[1]{{\color{blue}\bf{BH: #1}\normalfont}}
% \newcommand {\jeremy}[1]{{\color{teal}\bf{JW: #1}\normalfont}}
% \newcommand {\fred}[1]{{\color{darkpurple}\bf{KK: #1}\normalfont}}

% comment out the above and uncomment these for final submit
\newcommand {\bjoern}[1]{ }
\newcommand {\jeremy}[1]{ }
\newcommand {\fred}[1]{ }

% command for adding new content
% \newcommand {\add}[1]{{\color{darkgreen}#1\normalfont}}
\newcommand {\add}[1]{{#1}} % for publication

\newcommand*{\quoted}[1]{{\small{\fontfamily{cmss}\selectfont{#1}}}}
\newcommand*{\participant}[1]{{\textbf{\small{\fontfamily{cmss}\selectfont{#1}}}}}
\newcommand{\etal}{et al. }

\newcommand {\bt}[1]{\textbf{#1} \normalfont}
\newcommand{\squishlist}{
 \begin{list}{$\bullet$}
  { \setlength{\itemsep}{0pt}
     \setlength{\parsep}{3pt}
     \setlength{\topsep}{3pt}
     \setlength{\partopsep}{0pt}
     \setlength{\leftmargin}{1.5em}
     \setlength{\labelwidth}{1em}
     \setlength{\labelsep}{0.5em} } }
\newcommand{\squishend}{
  \end{list}  }

\graphicspath{{./figures}}
\title{Interactive Flexible Style Transfer for Vector Graphics}

\author{Jeremy Warner}
\affiliation{
  \institution{UC Berkeley}
  \city{Berkeley, CA}
  \country{USA}
}
\author{Kyu Won Kim}
\affiliation{
  \institution{UC Berkeley}
  \city{Berkeley, CA}
  \country{USA}
}
\author{Björn Hartmann}
\affiliation{
  \institution{UC Berkeley}
  \city{Berkeley, CA}
  \country{USA}
}

\renewcommand{\shortauthors}{Warner, et al.}

\begin{abstract}

Vector graphics are an industry-standard way to represent and share visual designs.
Designers frequently source and incorporate styles from existing designs into their work. 
Unfortunately, popular design tools are not well suited for this task.
We present VST, \emph{Vector Style Transfer}, a novel design tool for flexibly transferring visual styles between vector graphics.
The core of VST lies in leveraging automation while respecting designers' tastes and the subjectivity inherent to style transfer.
In VST, designers tune a cross-design element correspondence and customize which style attributes to change.
We report results from a user study in which designers used VST to control style transfer between several designs, including designs participants created with external tools beforehand.
VST shows that enabling design correspondence tuning and customization is one way to support interactive, flexible style transfer.

\end{abstract}



\begin{teaserfigure}
    \centering
  \includegraphics[width=\textwidth]{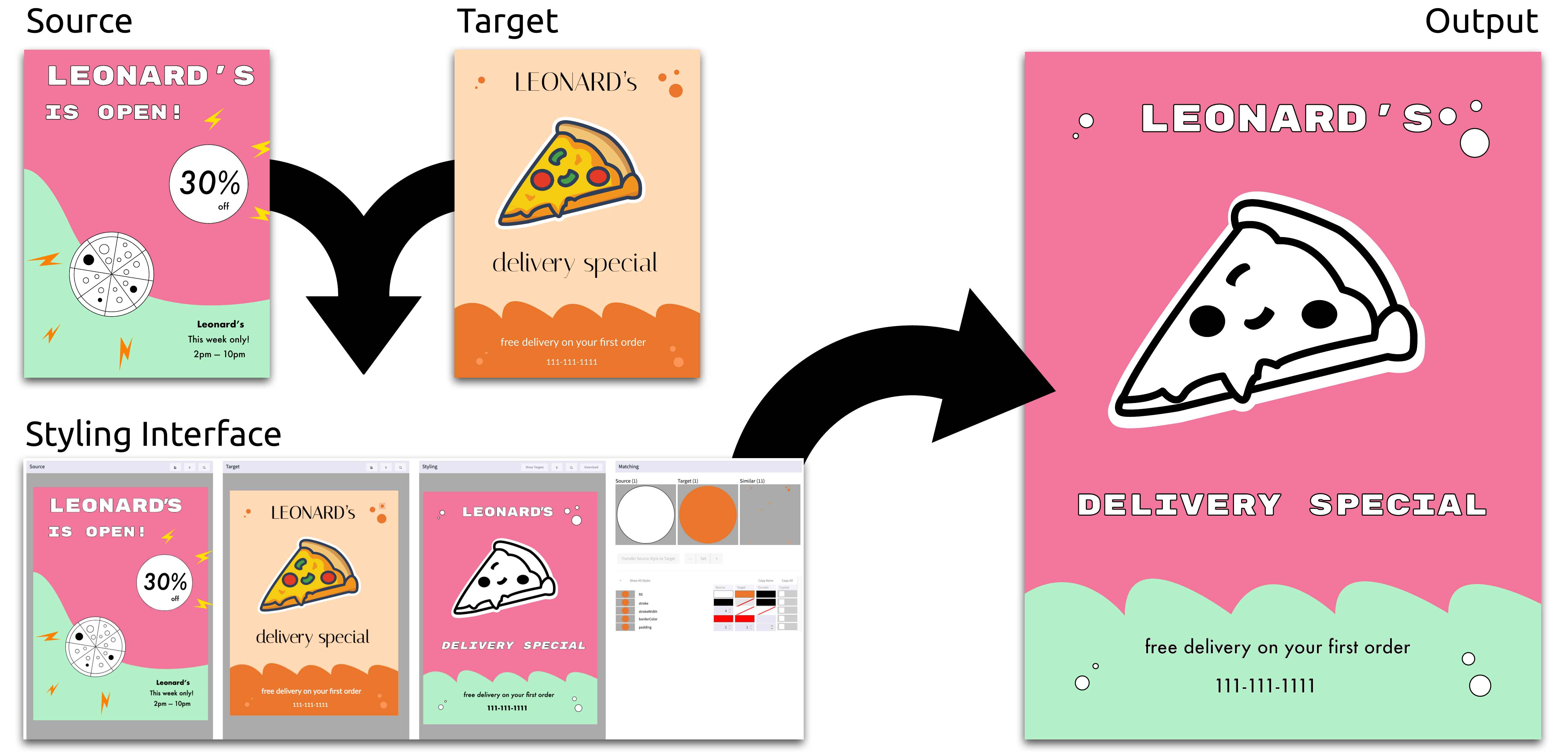}
  \Description[Style Transfer Overview and Teaser Figure]{VST generates new output graphics by transferring visual styles from source graphics onto target graphics. The output graphics retain a similar structure to the original target while bearing visual styles from the source graphics.}
  \caption{
  VST generates new output graphics by transferring visual styles from source graphics onto target graphics.
  The styling interface lets designers customize which styles to transfer, filter which elements to stylize, and preview the new stylized output graphics.
  The output graphics retain a similar structure to the target graphics while bearing styles from the source graphics.
  }
  \label{fig:teaser}
\end{teaserfigure}
\maketitle

\section{Introduction}

Vector graphics are an industry-standard way to represent and share a broad range of designs.
As a design medium, vector graphics offer compelling advantages, including scalability and precision.
Vector graphic designs store information about each graphical element that they contain.
This information enables editing the design at a higher level of semantics when compared to pixels.
Many vector graphics design tools have achieved success supporting designers working in this medium (e.g., Adobe Illustrator, Figma, Canva, Sketch).

Designers often edit vector graphics' overall appearance or style while retaining their underlying content and structure.
In this work, when we write \emph{style}, we refer to the defining visual properties of a design's elements (e.g., color, shape, size, and font).
Many alternative and valid definitions of this broad term exist.
Style editing tasks arise in multiple situations, such as applying inspirations from a mood board, updating existing graphics to a new visual identity, or exploring multiple alternative style variations.
For example, both a novice designer seeking to apply styles from a more polished design to their work and an experienced designer creating several variations of a similar design to present to a client for feedback face this task.
This complex task requires many selection and editing operations for different groups of objects.
Updating a design to conform to a new visual style can be exceptionally tedious and limits the exploration of different styles, even for experienced designers.

One potential solution is to use document-level themes or rules that consistently apply visual attributes to classes of objects.
This approach is standard across many design and presentation software tools.
For example, web pages use CSS (Cascading Style Sheets) to enable document-level styling, but these style-content links must be manually created and maintained.
A notable downside of using document themes or stylesheets is their \emph{rigidity}.
Compelling themes require element class information and pre-planning, introducing \emph{viscosity}~\cite{green1996usability} into the authoring process.
Despite CSS support in SVG \cite{svgSpec} via the \texttt{<use>} tag \cite{svgUse}, most vector graphics avoid it. 

Another promising direction is to \emph{automatically} transfer visual styles between graphics using information on how two given designs relate to each other.
However, this approach often fails to transfer styles as each designer uniquely intends.
This failure stems from two sources: 1) the accuracy limitations of the algorithm and 2) the inherent subjectivity around \textit{good} style and varying tastes that designers may have.
A fully automated approach may transfer styles in undesired or unpredictable ways. 
The lack of adequate designer controls is a clear barrier to levering automation \cite{automation2019Vogel}.

A tool should enable rapid iterating on different possible style transfer results to address the shortcomings of a fully automatic style transfer approach.
Our research aims to combine the benefits of automation with effective controls for customizing and exploring design variations.,
Our approach combines automatically generated design correspondences with interactive control of how and where to transfer styles.
We leverage prior work \cite{Shin2021MultilevelCV} on generating an automatic correspondence between vector graphics.
This method yields a between-design element correspondence (Fig.~\ref{fig:element-relation}) and element-wise similarity along multiple dimensions.

We present a new design tool, VST, short for \emph{Vector Style Transfer}.
VST provides designers with an interface to visualize and customize how style flows across designs (Fig.~\ref{fig:workflow}).
VST displays a dynamic list of element styles, allowing designers to easily copy, reset, and customize element style attributes (see Appendix~\ref{svg-properties} for all attributes).
With VST, designers can map and remap example \textsc{Source} element styles onto contextually similar elements. 
VST also features fast and flexible ways to identify, select, and style \textsc{Target} elements.
The \textsc{Output} canvas re-renders the stylized \textsc{Target} graphics in real-time with any changes, providing immediate visual feedback.

Conceptually, VST expands the \textit{eyedropper} or element-wise \textit{style copy-paste} interactions to groups of elements.
VST can infer many element relations directly, omitting the need for explicit element structure or class information.
Our combined automation-powered interactive style transfer approach means that designers can get the best of both worlds --
their style definitions can both be based on ad-hoc demonstrations and quick to apply flexibly across designs.

To evaluate VST's style transfer capability, we recruited six designers to transfer styles between nine designs.
Each designer participating in the study successfully used VST to interactively transfer styles to their satisfaction and make nine new \textsc{Output} designs.
In a follow-up design replication study, we recruited four expert designers to each manually replicate six of these \textsc{Output} designs in their preferred design tool.
\add{The results from this preliminary study suggest that someone using VST may reduce the time and work for this style transfer task compared to experienced designers using industry-standard tools.}
Our contributions include the following:

\begin{enumerate}
    \item VST, a design tool that introduces a novel user interface for interactive, user-guided, flexible style transfer for vector graphics. Its key interaction principles are: a) enabling users to edit computed correspondences at multiple levels, and b) enabling users to customize how attributes are transferred between designs across the correspondence.
    \item Two user studies that demonstrate: a) that designers can successfully transfer styles between graphics with VST, and b) that designers without VST can spend more time and effort to produce equivalent design results.
\end{enumerate}

\section{Related Work}

The most relevant prior work follows several themes: supporting creative processes with automation, inferring design structures, automatic transfer techniques, and other advanced vector graphics design tools.
We review each of these in turn.

\subsection{Supporting Creative Processes with AI}

While automation is powerful, gracefully integrating it into existing creative practices demands care.
Regarding working with AI as a design material, scholars have elaborated on the need for retaining control \cite{automation2019Vogel, yildirim2022HowEd, suh2021socialglue, palani2022factory, davis2021cocreative, roy2019automation}.
For GUI design, Dayama et al. present a method for interactive layout transfer, where the layout of a source design is transferred automatically using a selected template layout while complying with relevant guidelines \cite{Dayama2021InteractiveLT}.
In photography, researchers have provided mechanisms for guiding photographers to optimize image aesthetics \cite{e2021dynamic} and to find ideal portrait lighting conditions \cite{e2019optimizing}.
Goal-oriented transformations can also be applied to existing designs (e.g., improving accessibility) \cite{Zhang2021ScreenRC} or to produce alternative designs for different viewports  \cite{Hoffswell2020TechniquesFF}.

Our rationale for using element relationships between designs as a primary mechanism for transfer is that this mirrors how designers tend to work already when manually transferring styles.
Highly related to our line of work are feedforward and example-driven corrections.
Feedforward work refers to showing the user the output or result of their action before it happens--a preview of applying different interface actions \cite{vermeulen2013normanBridge, djajadiningrat2002but, victor2012affordance}.
For example, OctoPocus provides dynamic guidance to bolster users' ability to learn stroke-based gestures \cite{octopocus}.
Example-driven corrections and interaction models like those in FlashMeta \cite{Polozov2015FlashMetaAF} or programming-by-demonstration disambiguation models \cite{UI_PBE} provide alternative techniques that address similar problems.
Feedforward and inherent feedback can promote UI element functionality understanding to users, though computing this information fast enough for live, interactive contexts can be challenging.
With that said, cluing in authors on their actions' impact is valuable.
For example, the Lightspeed rendering pipeline enabled interactive prototyping of professional 3D graphics, enabling more design variation exploration \cite{ragan2007lightspeed}.
One approach might leverage lower-fidelity previews of variations when interacting with automation, such as design galleries.
We avoid using design galleries as our early prototypes showed the varying complexity and breadth were visually overwhelming.
For an analogy in text editing: VST spell-checks the entire document, while feedforward suggests autocompletion options given what is already written.

Example-based corrections generate a program that satisfies all demonstrated changes, iteratively growing more complex.
Example-based style retargeting for websites provides a successful analog to vector graphic style transfer in HTML/CSS \cite{Kumar2011BricolageER, benson2013cts}.
Example galleries can effectively support open-ended design authoring, where styles come from potentially multiple sources \cite{Lee2010DesigningWI}.
While the document-object-model hierarchy is essential to styling web pages, such grouping structures and labels are entirely optional and often absent in vector graphics.
Groups may be constructed arbitrarily (e.g., for editing convenience) rather than having any consistent semantic meaning.
Designers can encode hierarchical information through groups but frequently opt to style elements directly \cite{roy2019automation}.
Bringing interactive style transfer to vector graphics is a unique problem.

\subsection{Inferring Design Structures}

\begin{figure*}[t]
  \includegraphics[width=1\textwidth]{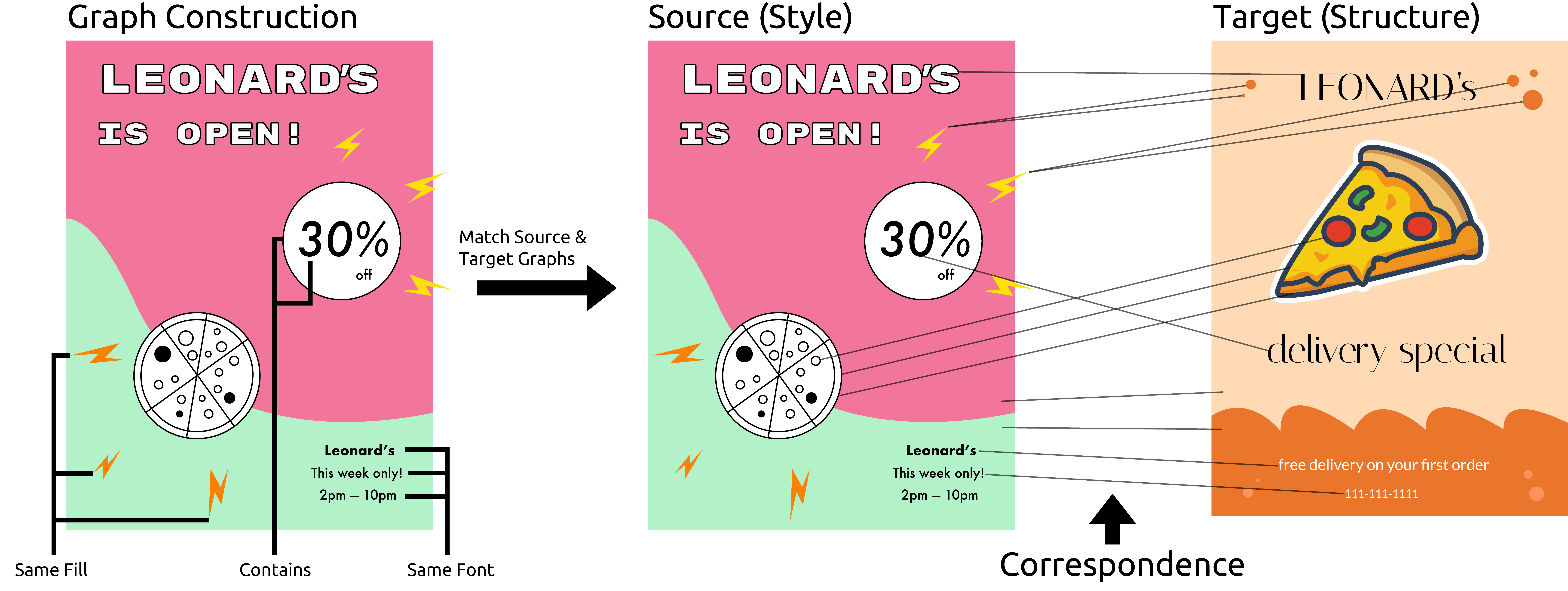}
  \Description[Graph construction and matching]{Vertices are primitive design elements (e.g., shapes, text, images), and edges are semantic relationships (e.g., same fill, containment, same font). We compute a correspondence between the two sets of design elements to enable styling across designs.}
  \caption{
  An overview of automated design correspondence.
  To relate design elements, we first construct a graph from each given design, where the \emph{vertices} are primitive design elements (e.g., shapes, text, images) and \emph{edges} are semantic relationships (e.g., same fill, containment, same font).
  Once the \textsc{Source} and \textsc{Target} graphs are constructed, we then compute a correspondence between the two designs' elements using the technique previously detailed in \cite{Shin2021MultilevelCV}.
  This automatically generated correspondence is VST's basis for (a)how to find similar elements \emph{within} a design (e.g., for easier selection/styling) and (b) identifying which elements are similar to each other \emph{across} designs (e.g., determining which initial styles to transfer).
  Each \textsc{Target} element is linked to a single \textsc{Source} element.
  Only a subset of links between these designs' elements are shown.
  }
  \label{fig:element-relation}
\end{figure*}

Researchers have used several approaches to infer underlying or implicit structures in visual designs.
Traditionally, this work primarily operates on some structured representation (like HTML or SVG).
For user interfaces, large libraries have helped to characterize and infer document structure \cite{Deka2017RicoAM, Kumar2013WebzeitgeistDM}.
Linking styles via direct manipulation and element cloning provide a clear view and control of an element's style properties \cite{Hoarau2012AugmentingTS}.
There is also work to recognize higher-level design patterns through designs by inducting grammars~\cite{Talton2012LearningDP}.
For the domain of D3 visualizations, Hoque et al. map data types onto shapes/axes to help search for relevant designs \cite{Hoque2020SearchingTV}.
Harper et al. showcase tools for deconstructing and restyling a D3 visualization by extracting the data and modifying visual attributes of marks \cite{Harper2014DeconstructingAR}.
More recent work also focuses on inferring design structure from images directly.
Computer vision techniques are improving on reverse engineering user interface models directly from screenshots \cite{Wu2021ScreenPT, Seelaudom2017ASF, Feiz2022UnderstandingSR}.
Similar work using vision-based methods has helped leverage attention towards answering questions and understanding mobile UIs \cite{li2023spotlight, swearngin2019modeling, schoop2022tap}.
Reddy et al. use differentiable compositing to identify pattern instances within a design \cite{Reddy2020DiscoveringPS}.
Scene graphs have also characterized structural relationships within and between 3D environments \cite{Fisher2011CharacterizingSR}.
For vector graphics, Shin et al. demonstrate a technique using graph kernels to find relationships between elements of designs \cite{Shin2021MultilevelCV}.
We leverage this preexisting automatic technique to compute a correspondence between design elements (like those shown in Fig.~\ref{fig:element-relation}).
The contribution of this work centers on our novel design tool that goes beyond pure algorithmic automation by enabling flexible interactions between the capabilities of such an algorithm and the designer's high-level styling goals.

\subsection{Automatic Transfer Techniques}

While automatic style transfer techniques can generate impressive image transformations, they are generally functional as \textit{theme selections}.
Due to the broad range of shape primitives, graphic designs do not immediately lend themselves to this document-level style transfer approach.
The selective extraction and transfer of specific styles are too precise to be encoded in a one-dimensional slider \cite{Isola2017ImagetoImageTW, Johnson2016PerceptualLF}.
The variations of vector designs also make mapping onto an otherwise standard template difficult (e.g., facial key points) \cite{Thies2019Face2FaceRF}.
Additionally, text can be used to edit image content and style directly \cite{brooks2022instructpix2pix}.
While layout is not our tool's focus, prior work highlights optimization techniques that can be used to automatically format text documents \cite{Hurst2009ReviewOA}.
ImagineNet restyles mobile apps with neural style transfer and updating assets in place \cite{imagineNet}.
To be stylized with image-based techniques, vector graphics must first be rasterized, losing future object-level awareness and scaling abilities.
The state of the art in automatic vector generation includes leveraging pixel-based diffusion models \cite{Jain2022VectorFusionTB} by leveraging a differentiable vector graphics representation \cite{Li:2020:DiffVG}.
DeepSVG uses GANs to generate and interpolate between SVG icons and shares a large-scale SVG dataset \cite{deepsvg}.
Kotovenko et al. model a painting using discrete strokes to recreate style transfer better \cite{Kotovenko2021RethinkingST}.
Within font, some work shows the possibility of even inferring and transferring style between font glyphs \cite{ Dhanuka2019VectorBG, Lopes2019ALR}.
These techniques often give users little to no control of \textit{how} the style is transferred.
Our work focuses on optimizing the potential value that these automatic approaches can provide by introducing meaningful high-leverage interactions to customize and control generated output while retaining the core vector graphics representation that designers are familiar with working with.

\subsection{Vector Graphics Design Tools}

Several techniques for authoring or adjusting vector graphics exist and inform this work.
Object-Oriented Drawing introduces a new way to create and style elements directly on the canvas \cite{Xia2016ObjectOrientedD}.
DataInk supports cloning and binding user-generated symbols to data, facilitating lightweight restyling \cite{Xia2018DataInkDA}.
Sketch-n-Sketch links drawing code and vector graphics, letting users directly edit the SVG in a canvas, modifying the code which generates it \cite{Hempel2019SketchnSketchOP}.
For mathematical diagramming, Penrose uses layout energy-minimization techniques coupled with a language for specifying explicit styles and content of what to render \cite{Ye2020PenroseFM}.
Falx uses user demonstrations and program synthesis to create new visualizations \cite{Wang2021FalxSV}.
Existing tools can even convert web designs into a vector layout \cite{html.to.design}.
Para supports binding procedural art generation constraints with graphics, including cases where there are many-to-many constraints \cite{Jacobs2017SupportingEP}.
A follow-up project, Dynamic Brushes, combined procedural programming into brush behavior and design, enabling more custom expression \cite{Jacobs2018ExtendingMD}.
Other design tools have looked at supporting design layout \cite{kikuchi2021modeling, ODonovan2015DesignScapeDW}, fashion \cite{Vasileva2018LearningTE}
and design coloring \cite{zhao2021selective, hegemann2023cocolor}.

\section{Vector Style Transfer}
When transferring styles between vector graphics, designers may identify an inspirational style they want to copy from a \textsc{Source} design.
Next, in a \textsc{Target} design, they may identify design elements they would like to stylize.
Then, they will update the stylistic attributes of those relevant \textsc{Target} elements using the \textsc{Source} style as a reference.
Alternatively, they may first focus on the \textsc{Target} design they wish to change and pull stylistic influences in from a range of \textsc{Sources}, exploring possible variations.
Generally, this styling is an iterative and flexible process that involves reasoning about (a) \textit{which elements correspond to each other across designs} and (b) \textit{which style attributes to transfer}.
There is subjectivity regarding the most desired application of style, and higher-level considerations like the overall cohesion of the \textsc{Target} design after styles have transferred further complicate this task.
The resulting \textsc{Output} design has the \textit{style} of one design and the \textit{content/structure} of another -- though this distinction is still inherently subjective. 
Still, this task (using examples to update existing graphics with new visual styles) is expected in the graphic design process \cite{lee2010designing, herring2009getting, koch2018surfing}.

\begin{figure*}[t]
    \centering
    \includegraphics[width=\textwidth]{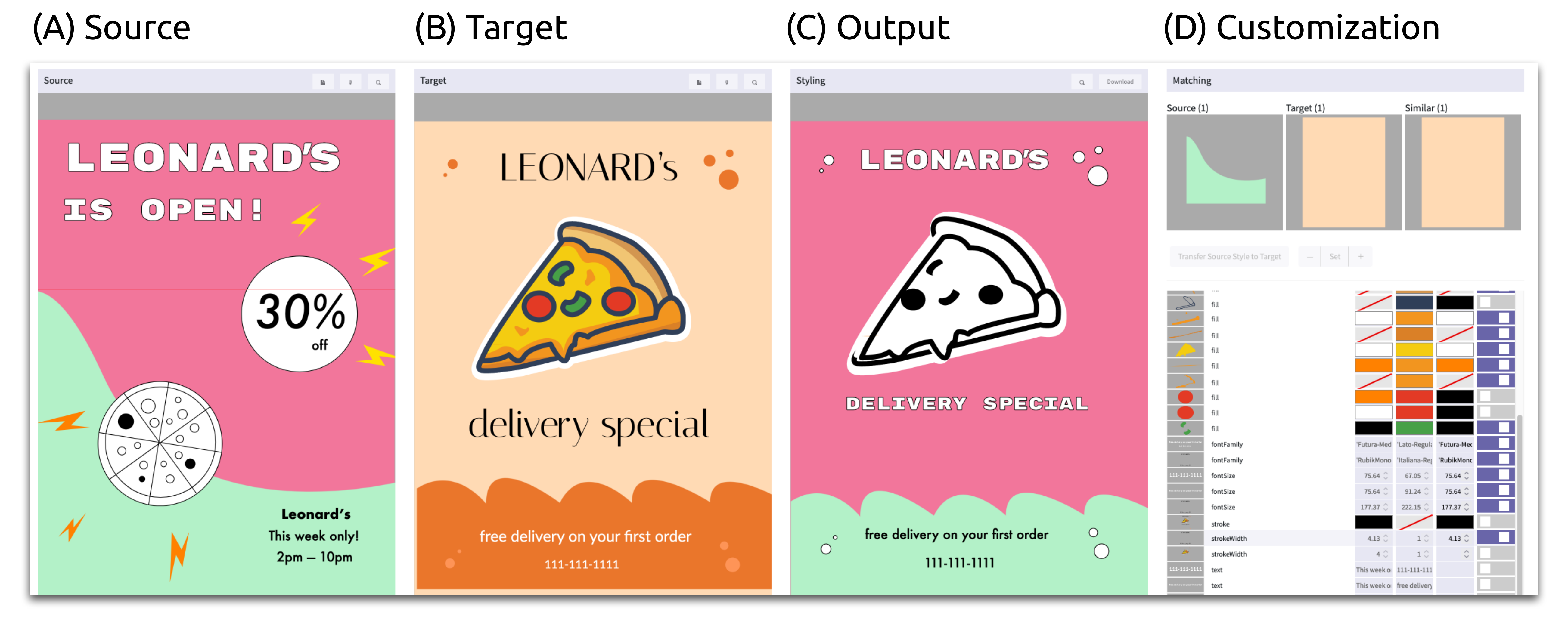}
    \Description[The VST Interface overview]{The VST Interface: (A) the Source graphics (where the style is sourced from), (B) the Target graphics, (C) the Output canvas (the current style transfer result), and (D) customization controls.}
    \caption{An overview of the VST interface, including (A) the \textsc{Source} graphics (where the style is sourced from), (B) the \textsc{Target} graphics, (C) the \textsc{Output} canvas (the current style transfer result), and (D) customization controls for matching element styles across canvases and filtering which style attributes to copy or modify.
    Designers can filter this list of attributes (shown in D) based on the current selection to do more focused editing or instead modify shared style attributes across the entire design.
  }
  \label{fig:workflow}
\end{figure*}

\subsection{Design Goals}
A high-quality element correspondence is one way to enable fast and effective style transfer for vector graphics designs.
To provide designers with flexible control over style transfer is to provide them with tools to control the correspondence between designs.
Moreover, to be worthwhile, the resulting designs should be of satisfying quality and faster to generate than existing tools, especially when considering the cost of learning to use a new tool.
Grounded in our literature review and personal experience editing graphics, we created these design goals for Vector Style Transfer (VST):

\begin{description}
     \item[DG1] Let designers powerfully tune design correspondences.
     \item[DG2] Enable flexible control over which styles are transferred.
     \item[DG3] Reduce the work and time needed for transferring styles.
 \end{description}
Our vision for how the functionality of VST best fits into existing processes is as a plugin or new tool in existing vector graphics design software.
Designers could select an object group and copy their style.
Then, they could select any other group within their design document and apply that style -- without manually selecting each element subset.
Additionally, they could filter which styling attributes they would like to copy.
This work could either be used as a starting point to render a design in several alternative styles or to make a set of designs adhere to a single style.

\subsection{Exemplar Scenario}
We will demonstrate VST's functionality with an exemplar scenario involving vector style transfer.
Consider Xavier, a designer hired by a local Italian restaurant, \textit{Leonard's}.
After a recent renovation, the restaurant is set to have a grand re-opening.
Xavier has created a new flyer to help them advertise, which the business manager approves.
To unify the brand's style, the business manager also asks him to create new versions of several existing graphics, including menus and a special delivery advertisement.
These designs should look like they all refer to the same restaurant.

This style unification process Xavier faces involves many repeated manual edits and cross-references.
Instead of manually ensuring exact visual consistency, he opens VST and loads in both graphics (\textsc{Source}: the new flyer, \textsc{Target}: the previous advertisement).
VST computes a correspondence between elements of these two designs and automatically copies styles between matches.
This correspondence technique ensures a one-to-many mapping from the \textsc{Source} elements to the \textsc{Target} elements.
This ensures that every \textsc{Target} element will be matched, while some \textsc{Source} elements may not be initially matched.
Xavier then sees the \textsc{Output} canvas update with newly stylized graphics (Fig.~\ref{fig:workflow}).

\begin{figure}[t]
        \centering
        \includegraphics[width=1\columnwidth]{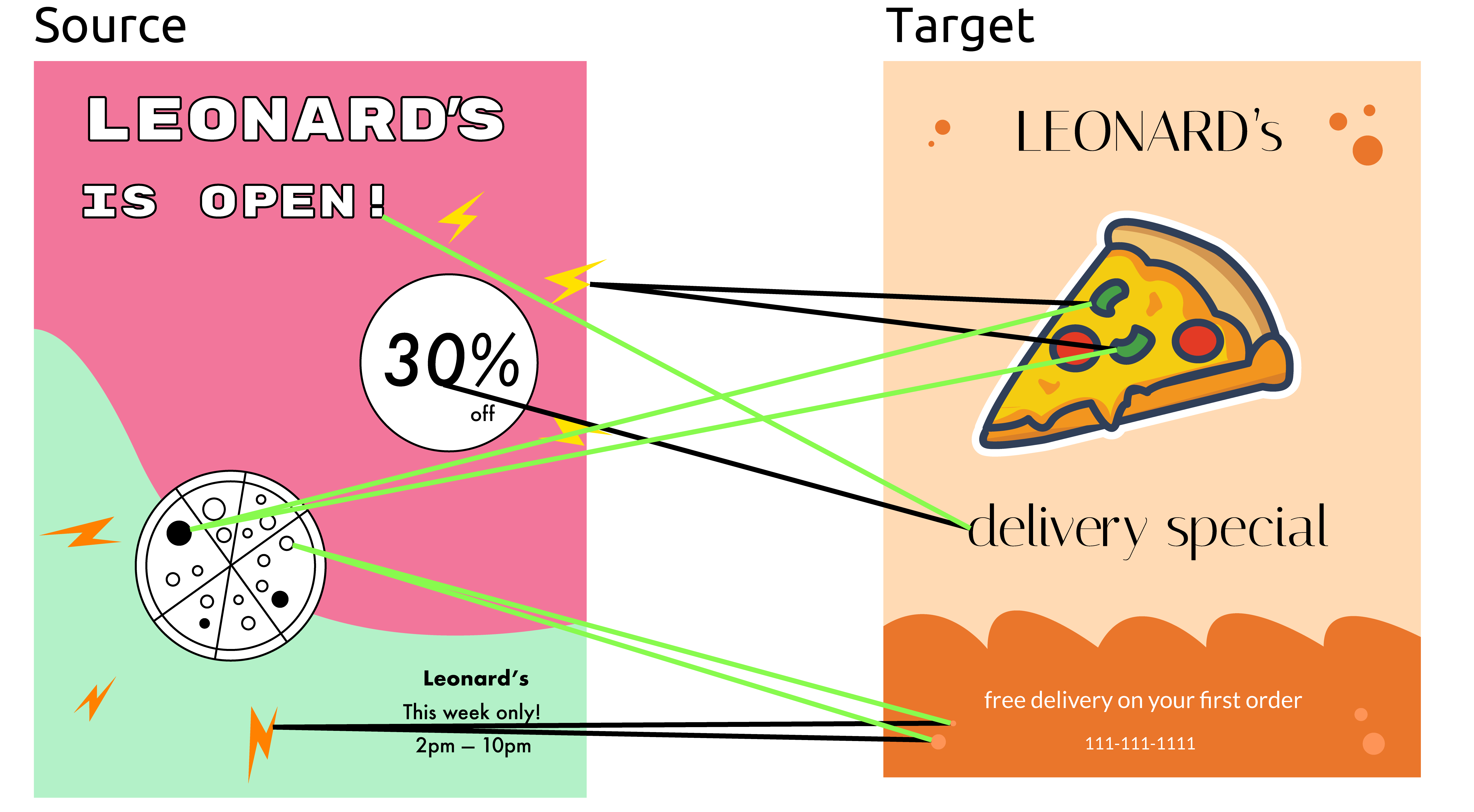}
        \Description[Desired and initial correspondence]{Desired correspondences black bars between designs connect elements initially, while green bars show a more desired match.}
        \caption{
        The black lines show an initial correspondence between the elements of the \textsc{Source} and \textsc{Target} designs.
        The green lines show an alternative, more desired set of links.
        When users select their desired \textsc{Source} and \textsc{Target} elements and press  \textit{Transfer Source Style}, VST will update these links, redirecting the flow of visual styles across designs. 
        }
        \label{fig:matching}
\end{figure}

For each \textsc{Target} element, styles are copied from the most similar \textsc{Source} element as determined by the design correspondence algorithm \cite{Shin2021MultilevelCV}.
In addition to seeing the updated target graphics, a list of changed style attributes is displayed on the right-hand side of the interface (Fig.~\ref{fig:workflow}D).
The breadth of style attributes and the range of possible valid matches between elements makes using a fully automatic approach difficult.
The inherent subjectivity of style also means this first attempt will not always be correct, especially for more complex and open-ended designs.
Xavier immediately detects outlier text elements that are visually misaligned with the \textsc{Source} style directly on the \textsc{Output} canvas (Fig.~\ref{fig:workflow}).
Designers are trained to use gestalt principles of perception to organize a design.
Incorrect style transfer will lead to visual violations of these principles, which are often easy to detect \cite{Mullet1994DesigningVI}.
This means that some elements likely have been `mismatched' by the correspondence algorithm (Fig.~\ref{fig:matching}).
Using the \textsc{Source} canvas (Fig.~\ref{fig:workflow}A), Xavier can then specify which \textsc{Source} element the incorrectly styled text fields should visually match.
When he presses the \textit{Transfer Source Style to Target} button (Fig.~\ref{fig:system-match}), VST renders styles from the \textsc{Source} element onto the \textsc{Target} selection in the rightmost \textsc{Output} canvas (Fig.~\ref{fig:workflow}C).
Behind the scenes, VST applies these fixes to a copy of the original correspondence, avoiding recomputing the entire correspondence after updates.

Still, manually selecting each target element to update is tedious.
To enable faster transfer, designers can double-click on any \textsc{Target} element to select similar elements, as determined by the design correspondence.
Repeatedly double-clicking an element iteratively grows the set of selected \textsc{Target} elements.
This feature mirrors the multi-click selection in other media, like toggling between word-sentence-paragraph selections within a text document.
Here, we use the underlying within-document element-wise similarity score to intelligently add elements most similar to the currently active selection.
A similarity score is computed for each element relative to the currently selected elements, and the elements with the highest score is added to the active selection.
Double-clicking on a \textsc{Source} element conversely selects all \textsc{Target} elements currently matched to that element, which shows how style flows from the \textsc{Source} to \textsc{Target} design.
The customization panel shows a pane of similar elements, where Xavier can preview this selection (Fig.~\ref{fig:system-match}).
Despite Xavier updating the \textsc{Source}-\textsc{Target} correspondence, the resulting \textsc{Output} design still has some problems.
For example, while the font and color are corrected, the copied font size makes some elements not fit neatly in the new design (Fig.~\ref{fig:sys-style}).
Once matched, VST has controls for customizing which specific style attributes are transferred.
To focus on the desired element, he clicks \textit{Show Filtered Style} to only see the styling applied to the text element (Fig.~\ref{fig:sys-style}).
He toggles the \texttt{fontSize} attribute, resetting that element's font size and updating the \textsc{Output} canvas.
Similar attribute values are grouped in this view to make selecting and editing easier.
He continues this style transfer process until he is satisfied with the quality of the new design.
Internally, these changes build up a list of attribute transformations to apply to the \textsc{Target} design.
The customization pane can highlight just the modified attributes, summarizing stylistic changes at a glance.
Finally, Xavier downloads the \textsc{Output} graphics from VST as an SVG file to save his work.

\subsection{Implementation}

We used ReactJS to build the VST interface and deployed our prototype online.
Vector graphics are rendered using FabricJS, a vector graphics library leveraging the HTML5 canvas backbone.
SVG files, such as those exported from industry-standard design tools like Sketch, Figma, Canva, and Adobe Illustrator, can be directly imported.
Once VST has imported the input \textsc{Source} and \textsc{Target} graphics, we compute a correspondence between the two designs using a comparison technique introduced by Shin et al. \cite{Shin2021MultilevelCV}.
This technique represents each design as a multigraph (rather than a typical parent-child hierarchy tree) to support matching elements across a broader range of similar attributes.
Vertices are primitive design elements (e.g., shapes, text, images), and edges represent semantic relationships between elements (e.g., alignment, containment, same fill).
This correspondence contains per-element similarity scores across several dimensions (e.g., color, shape, size, and text).
In our implementation, correspondences between 20 or fewer elements are generally computed in real-time (< 1s).
Though slower, our study's larger design pairs are still tractable to match, with the largest pair (185 total elements) taking about 100s.
Our example set's average matching time per design pair (across Style Transfer Tasks 1 and 2) is $7.78$s.
Once obtained, match information can be exported and saved for later use.
\add{
A version of VST for styling pre-matched design pair examples is available at: \texttt{\url{https://berkeleyhci.github.io/vst/}}.
}

\section{Evaluations}

Style preferences are subjective, which means that making absolute statements about a style transfer tool's \textit{performance} is difficult.
Still, we sought to evaluate three key research questions:

 \begin{description}
     \item[RQ1] How would designers use VST for style transfer?
     \item[RQ2] Could VST stylize realistic, open-ended designs?
     \item[RQ3] Could VST reduce the time or work of styling?
 \end{description}

\subsection{Style Transfer Evaluation}

\textbf{Method} --
To answer RQ1 and RQ2, we ran an exploratory study with six experienced designers (\participant{D1-6}).
Before the study began, we asked designers to create a new design from a given prompt with their preferred design tool.
The prompt requested a single menu page design for a local restaurant's (\textit{Leonard's}) mobile phone application.
The goal was to include designer-provided source graphics to create a more realistic style transfer scenario.
More methodology details are available in Appendix~\ref{sec:methodology}, and more information about the participant's background is in Appendix~\ref{sec:participants}.

\textbf{Task 1: Basic Graphics Pairs} --
After an interface demo and the opportunity to ask questions, designers used VST to transfer styles between five pairs of example designs that the authors prepared. 
The design pairs we chose for designers to transfer from are shown in Fig.~\ref{fig:author-transfer} (T1.1-5).
We chose these graphics to capture a breadth of different graphic design domains (e.g., art, infographics, UI mockups).
We instructed designers to apply styles from the \textsc{Source} to the \textsc{Target} graphics to make the \textsc{Source} and \textsc{Output} as stylistically similar as possible.
Once satisfied, they would save the \textsc{Output} graphics and move on to the next pair.
 
\textbf{Task 2: Open-Ended Transfer} --
To observe how VST handled styling more open-ended realistic designs (RQ2), designers transferred styles from their externally created designs onto three new related templates (T2.1-3).
In these tasks, the \textsc{Source} was a menu page created by each designer before the study with their preferred design tool.
We matched their designs to three new template pages (a loading screen, a reviews page, and a checkout cart), all for \textit{Leonard's} mobile app.
The generated output design correspondences (Fig.~\ref{fig:element-relation}) were not hand-tuned at all before the study.

\subsection{Style Transfer Results}

\begin{figure}[t]
  \centering
  \includegraphics[width=1\columnwidth]{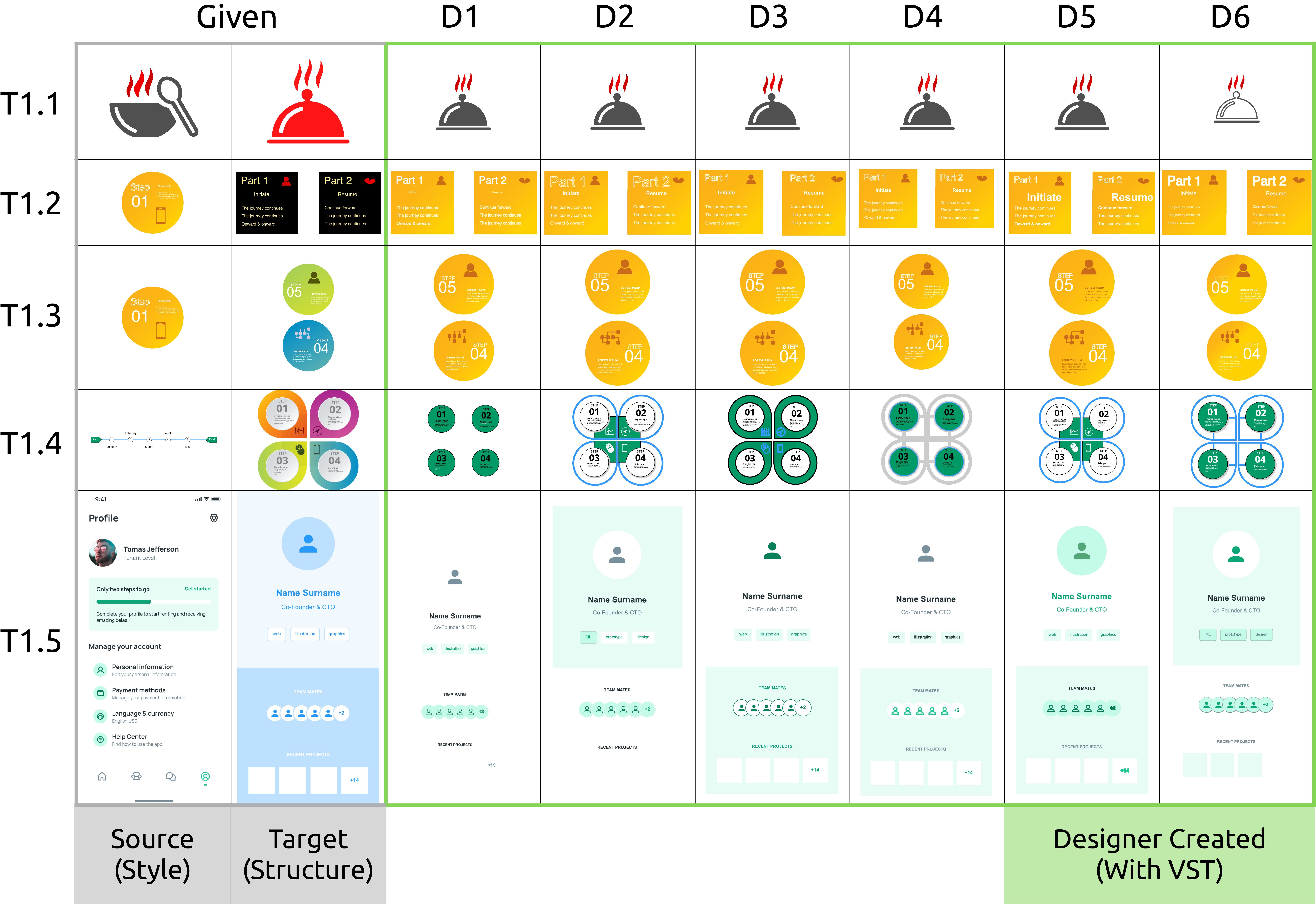}
 \Description[Design Task 1 of Style Transfer Evaluation]{Task 1: Recruited designers used VST to create new designs by applying styles from Source onto Target graphics. Both Source and Target designs were provided to the designer here.}
  \caption{
  Task 1 (Style Transfer) -- Basic Graphics Pairs.
  Here, designers \participant{D1-6} used VST to transfer styles from the \textsc{Source} to the \textsc{Target} graphics.
  Both \textsc{Source} and \textsc{Target} designs were provided to the designers.
  In simpler cases, the design transfer result is uniform across designers (T1.1-3).
  Still, despite each designer starting from the same pair of designs, variations arose in more complex design pairs (T1.4-5).
  }
  \label{fig:author-transfer}
\end{figure}

\begin{figure}[ht]
  \centering
  \includegraphics[width=\columnwidth]{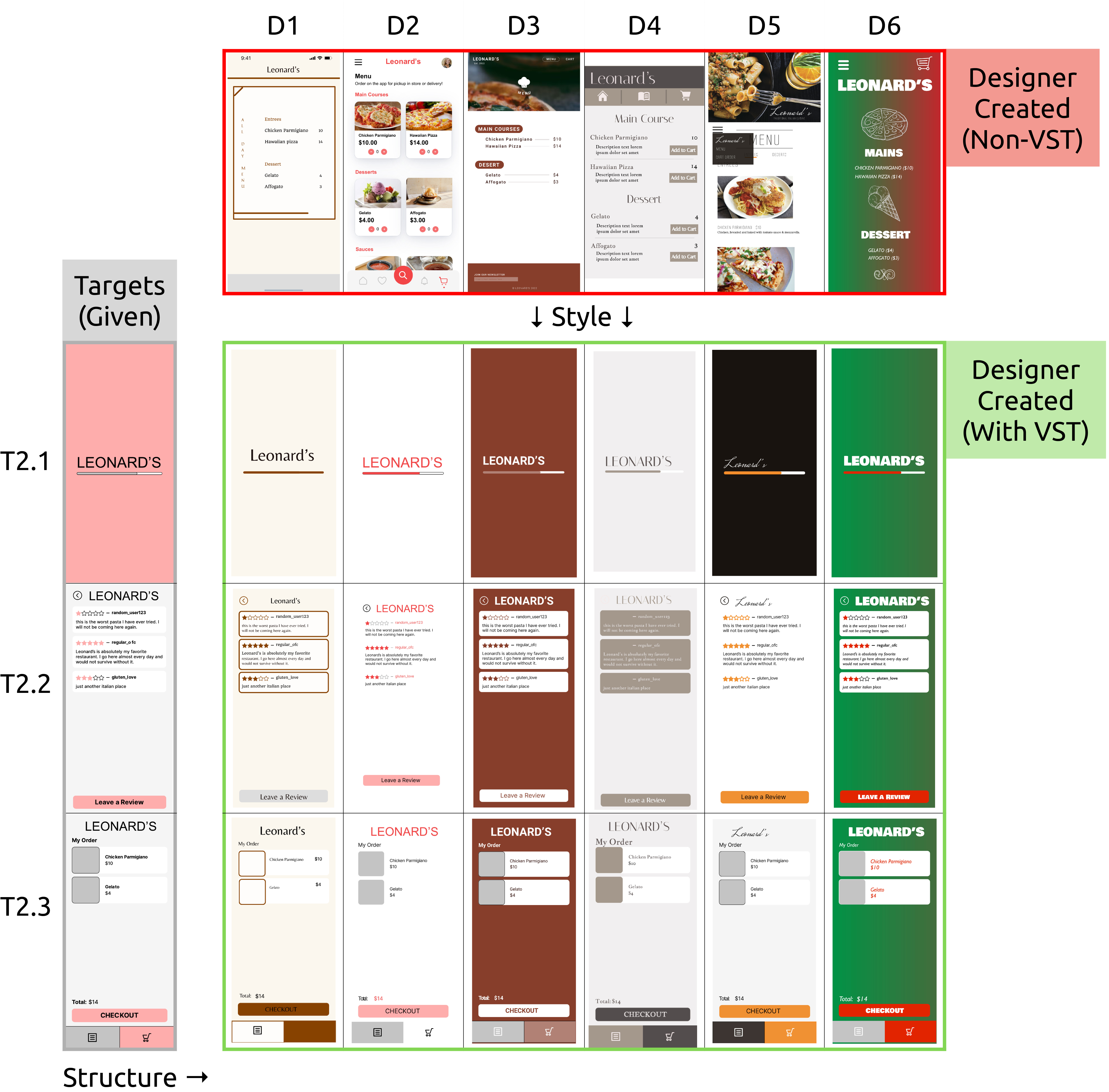}
 \Description[Design Task 2 of Style Transfer Evaluation]{Task 2: columns show unique designs that designers brought into the study, and rows show author-provided design templates. The inner table shows new designs created by applying styles from their own non-VST creation onto target templates. Inspecting each column shows a clear, unified visual style, while each row shows the structure of the Target.}
  \caption{
  Task 2 (Style Transfer) -- Open-Ended Transfer.
  Before the study, we gave designers (\participant{D1-6}) a prompt for a menu design with specific elements without any style instructions.
  The column header shows designs that they brought into the study (\textsc{Sources}), and the row header shows design templates (\textsc{Targets}).
  The inner table shows new designs created by applying styles from their externally created \textsc{Source} design onto previously unseen \textsc{Target} templates.
  Inspecting each column shows a unified visual style inherited from the \textsc{Source} document, while rows show the \textsc{Target} structure.
  }
  \label{fig:user-transfer}
\end{figure}

Our style transfer evaluation study found that designers could use VST to control style transfer across basic designs (RQ1), even generating variety in their \textsc{Output} designs from the same inputs.
Those designers successfully used VST to flexibly transfer styles from more realistic, open-ended designs created with external tools (RQ2).
We take this as an indication that VST enabled the style transfer it was designed to support.
Each designer participating in the study (D1-6) used VST to generate eight new \textsc{Output} designs successfully.
Designers also answered Likert-scale questions regarding their experience with VST (Fig.~\ref{fig:likert}).
Style transfer examples from the evaluation are shown in Figures~\ref{fig:author-transfer} and \ref{fig:user-transfer}.

Designers, despite never using a similar interface before, used VST's features to both (a) modify design correspondences (DG1) and (b) filter and edit styles per correspondence (DG2).
Software instrumentation revealed that almost all designers on almost all tasks used VST to tune computed correspondence matches.
On average, designers performed 6 such corrections per task.
While making these corrections, designers used the functionality to \textit{select similar} elements to the ones they manually selected.
On average, designers performed 7.3 similarity selections and spent about 4.8 minutes per task. 
As a reminder, designers were only instructed to match the styles to the best of their ability -- not to do so as quickly or efficiently as possible.
\add{
We showcase additional, more complex VST graphics made outside of this study in the Appendix (Fig.~\ref{fig:add-examples}) and in our paper's accompanying project video.
}

\textbf{VST let designers tune design correspondences (DG1).}
Overall, designers appreciated the style transfer control that VST provided them.
The designers' Likert-scale responses indicated they could produce designs they were satisfied with (Fig.~\ref{fig:likert}).
Most designers could see themselves using the tool again and found VST flexible enough to perform style transfer as they intended.
Their verbal remarks are corroborated by the frequency with which they used the correspondence correction feature (Average: $\mu: 6.0$, Standard Deviation: $\sigma = 3.8$) and attribute editing feature ($\mu: 24.0$, $\sigma = 17.3$).

\textbf{VST enabled flexible control of style transfer (DG2).}
The designers created a wide variety of designs, even when given the same input graphics (Fig.~\ref{fig:author-transfer}).
For their own provided graphics, designers reproduced a consistent theme across a set of provided vector graphics templates (Fig.~\ref{fig:user-transfer}).
Several designers remarked on the convenience of reusing visual styles directly.
\quoted{\participant{D4:} Very fun! Appealing to a visual thinker who values efficiency and hates repeatedly doing the same things. Magical, "it read my mind!" kind of feeling.}
While most found it clear how to use the different parts of the prototype to achieve their desired style transfer, there was also feedback that the transfer results were sometimes surprising.
This surprise likely stemmed from having multiple ways to style elements (e.g., tuning the correspondence vs. what styles the correspondence transfers).

\textbf{Designers enjoyed applying broad changes.}
Designers valued the ability to apply broad style changes quickly.
\quoted{\participant{D3}: I was impressed by how well the system generated its "best guess" when I selected the "Copy All." I also thought it was easy to learn and intuitive. It had tools that worked similarly to design software I already used (like dragging values to change the font size).}
\quoted{\participant{D5:} I liked how efficient the transferring process was in closely replicating the desired style with just a button. Even if it wasn't completely accurate, the toggle buttons under Copy All made fine-tuning specific aspects of design elements easy -- I could definitely see how this interface could reduce the amount of time that a designer would need to update designs.
}
Designers also appreciated directly selecting similar elements easily, which helped broader styling.
\quoted{\participant{D4:} Being able to select multiple elements precisely is very nice.}

\begin{figure*}[ht]
  \centering
  \includegraphics[width=\textwidth]{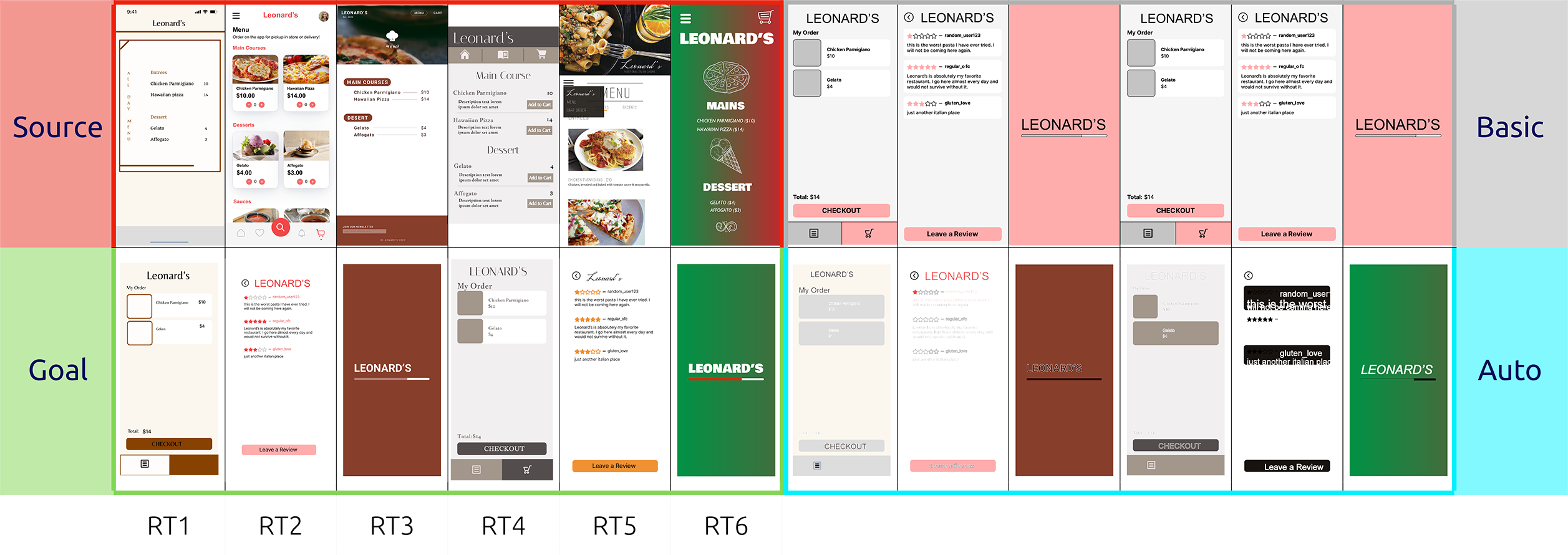}
 \Description[Design Replication]{A new set of expert designers (\participant{RD1-4}) used two different approaches to replicating six previous designs created (T2.1-3). Source styles and target structures are reused from the previous study and provided in vector format. The replication goal is provided as a reference image.}
  \caption{
Design Replication Task --
A new set of expert designers (replication designers \participant{RD1-4}) replicated six reference designs (RT1-6) from the previous style transfer evaluation tasks (Fig.~\ref{fig:user-transfer}) using two different starting points: \textit{Basic} and \textit{Auto}.
The first approach involved using Illustrator to transform the \textit{Basic} input design to the replication goal.
The second approach again used Illustrator, but instead has the algorithmic output (\textit{Auto}) as the starting point.
We provided the \participant{RD}s with source styles and target structures from the previous study in vector form and a reference image of the replication goal for both approaches.
  }
  \label{fig:replication}
\end{figure*}

\textbf{Correspondence-based transfer presents novel controls.}
No designer reported using a similar style transfer design tool before this study.
\quoted{\participant{D6:} I have not used anything that performed this exact function before, but I've used a tool to try to analyze an image and find out what fonts were used. It was not as reliable as this tool.}
While most designers (4/6) indicated an interest in using the tool again, others were hesitant, citing VST's deviation from the types of tools they were familiar with.
Some designers recognized the value of a style transfer tool:
\quoted{\participant{D4:} I have manually copied styles and have had other humans manually copy my own.
When successful, this tool manages to give you that feeling of empathy and creative connection (``Wow, the other designer understood my aesthetic and was able to replicate it! I feel they really understand my vision''). When it is not successful, it is easier and less stressful to correct than a human might be.
Plus, it is faster than asking another designer, fewer resources, less risk, and when it is successful, high reward!
}

\subsection{Design Replication Evaluation}

\textbf{Method} --
To answer RQ3, we ran a follow-up study.
Our goal with this study was to compare the time and work required for style transfer in VST with that of an expert using industry-standard design software.
We recruited four new expert designers as replication designers (\participant{RD1-4}).
More information about their background is in Appendix~\ref{sec:participants}.
They were tasked with recreating a subset of the \textsc{Output} graphics from the previous study (T2.1-3) in their preferred design tool (Adobe Illustrator).
Given that VST is a novel design tool there are no users with equivalent VST expertise comparable to the \participant{RD}s' Illustrator skill.
To approximate the performance of an expert VST user, the authors used VST to generate the same \textsc{Output} designs using the same input materials provided to the \participant{RD}s.
This data is labeled VST in Table~\ref{table:replication}.
Further methodology details are in Appendix~\ref{sec:methodology}.
We report the comparison between these three design replication methods in our results.

\textbf{Task: Design Replication} --
We selected six \textsc{Output} design examples from Fig.~\ref{fig:user-transfer} for this designer to replicate in Illustrator (\textit{Goal} in Fig.~\ref{fig:replication}).
We selected designs to include both graphics from every task (T2.1-3) that we gave the original designers and to include one example per designer (\participant{D1-6}).
We provided the \participant{RD}s with the \textsc{Source} and \textsc{Target} vector graphics files and an image of the generated \textsc{Output} (created initially by \participant{D1-6}).
The \participant{RD}s were then tasked with transforming the \textsc{Target} graphics to resemble the provided \textsc{Output}.
To measure what human adjustment is needed when working with the automatically stylized designs, we also asked the \participant{RD}s to replicate the \textsc{Output} starting with the initial automatically stylized \textsc{Output} graphics from VST.
These graphics (\textit{Auto}) are created by copying all styles using the initial automatic \textsc{Source} and \textsc{Target} correspondence.
We asked the \participant{RD}s to transform the now-partially stylized \textsc{Target} graphics to resemble the \textsc{Output} image.
Any difference between these two sets (\textit{Basic} and \textit{Auto}) would highlight the algorithm's impact on the task time and work.
To compare the potential of VST and existing tools, the authors also replicated the same \textsc{Output} designs from the previous study using VST (RT1-6).
The same input materials were used as in the Illustrator replication: the \textsc{Source} and \textsc{Target} vector graphics files and an \textsc{Output} image.

\subsection{Design Replication Results}

\begin{figure*}[ht]
    \centering
    \includegraphics[width=0.33\linewidth]{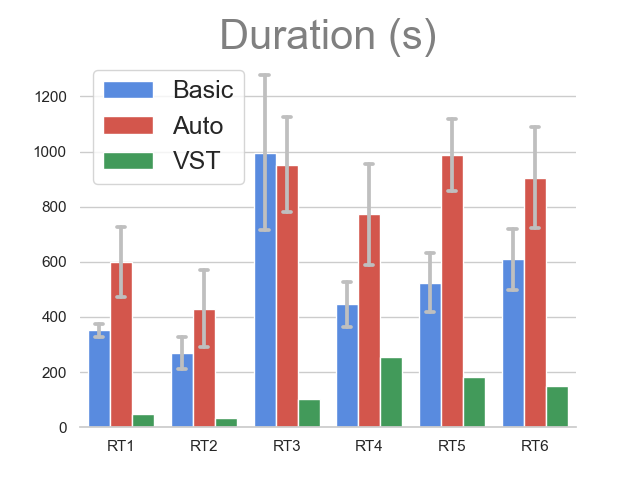}
    \hfill
    \includegraphics[width=0.33\linewidth]{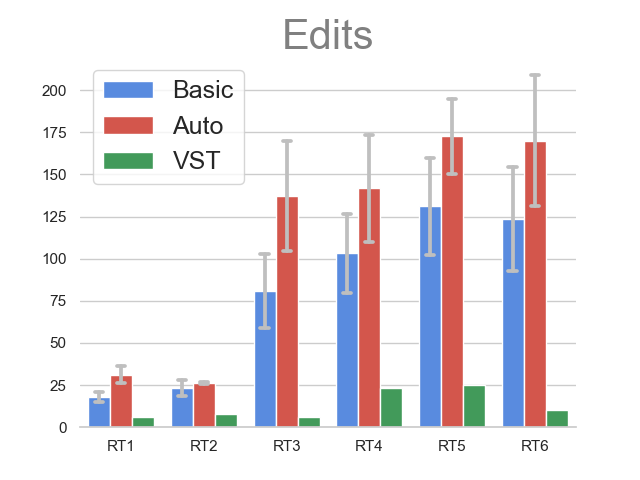}
    \hfill
    \includegraphics[width=0.33\linewidth]{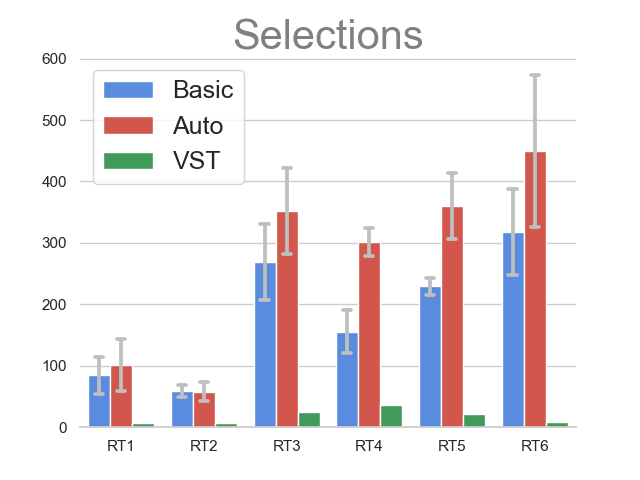}
    \Description[Follow-up Comparison]{Using VST was faster and required less edit and selection operations. Plots of duration, edits, and selections data from the design replication (RT1-6).}
    \caption{
Plots of the duration, edits, and selections data from the design replication (RT1-6).
Along each recorded measure (duration, edits, and selections), the authors using VST outperformed all four expert designers using Adobe Illustrator in replicating the stylized designs.
The \textit{Basic} and \textit{Auto} plots also include ticks showing the \textit{standard error} for each task computed over \participant{RD1-4}.
VST was only used once per task to obtain a baseline, so there are no comparable ticks to show.
    }
    \label{fig:rec-stats}
\end{figure*}

In our study, using VST to transfer styles was faster than expert replication designers (\participant{RD1-4}) transferring styles within their preferred design tool (RQ3).
The \participant{RD}s also performed more edit and selection operations using Illustrator than the authors using VST.
We report total work as a combination of selection and edit operations.
On average, the \participant{RD}s spent 534 seconds replicating from scratch (\textit{Basic}) and 774 seconds replicating from the output of the correspondence algorithm (\textit{Auto}).
In comparison, the authors required, on average, 129 seconds to match styles using VST.
A plot of the duration for each task is shown in Fig.~\ref{fig:rec-stats}.
Stats averaged over all tasks (RT1-6) are shown in Table~\ref{table:replication}.
Each replication designer also reported the style replication task as difficult and tedious.

\textbf{Transferring styles with existing tools is tedious.}
After replicating the designs in Fig.~\ref{fig:replication} (RT1-6), the \participant{RD}s reported on their experience by answering Likert-scale (ranging from 1-7) and open-ended survey questions.
They reported that using Illustrator for this style matching task is tedious for both starting points, with \textit{Auto} slightly more tedious than \textit{Basic} (Average ($\mu$): 6.8 $\rightarrow$ 5.8, Standard Deviation: $\sigma_{Basic} = 1.3$, $\sigma_{Auto} = 0.5$).
The associated scale labels were: 1-\textit{Not tedious at all} and 7-\textit{Extremely tedious}.
They also reported starting from \textit{Auto} was less fun than \textit{Basic} ($\mu$: 2.0 $\rightarrow 3.8$), with 1-\textit{Not fun at all} and 7-\textit{Extremely fun} ($\sigma_{Basic} = 1.0$, $\sigma_{Auto} = 0.8$).

\textbf{Editing from Auto was not faster than Basic.}
Combining automated style transfer with existing design software tools may even hinder designer performance.
The \participant{RD}s reached roughly the same Likert-scale level of satisfaction with  their final designs' quality from both the \textit{Basic} and \textit{Auto} starting points
($\mu_{Basic}$ = 4.3, $\mu_{Auto}$ = 4.5), with 1-\textit{Completely dissatisfied} and 7-\textit{Completely satisfied}
($\sigma_{Basic} = 1.0$, $\sigma_{Auto} = 1.0$).
However, they reported that generating the desired \textsc{Output} was harder with \textit{Auto} than \textit{Basic} ($\mu$: 6.3 $\rightarrow 5.0$), with 1-\textit{Not difficult at all} and 7-\textit{Extremely difficult} ($\sigma_{Basic} = 1.0$, $\sigma_{Auto} = 0.8$).
These stats match their written feedback:
\quoted{\participant{RD1:} Editing the auto files is harder -- there's more variance in the output, and sometimes unnecessary properties were added from the automatic transfer.}
\quoted{\participant{RD2:} In the standard [Basic] file, editing elements is more straightforward, while for the modified [Auto] one, I spent some extra time cleaning.}
\quoted{\participant{RD4:} I largely had a similar approach to both design files, though the original [Basic] one tended to be easier.}

\textbf{Replication designers wanted transfer tools like VST.}
After briefly interacting with VST at the end of the study, all \participant{RD}s were genuinely interested in trying out an Adobe Illustrator plugin with similar functionality ($\mu = 6.25, \sigma = 1.0$), with 1-\textit{Not at all interested} and 7-\textit{Extremely interested}.
\quoted{\participant{RD4:} The prototype looks very interesting!}
\quoted{\participant{RD1:} I would definitely try it when I want to apply vector-based styles to my design.}
When asked about if and where they would find VST useful:
\quoted{\participant{RD1:} I can see how this tool would be beneficial for tasks like redesigning an existing UI or early-stage exploration.}
When asked about other similar tools they have used:
\quoted{\participant{RD2:} In Figma, we save the font/color as a library preset, then when we change the setting, it automatically updates the components.}
\quoted{\participant{RD3:} The style transfer prototype is more adaptive than design components because files that I need to change may not have a component system.}

\begin{figure}[t]
  \centering
  \includegraphics[width=\columnwidth]{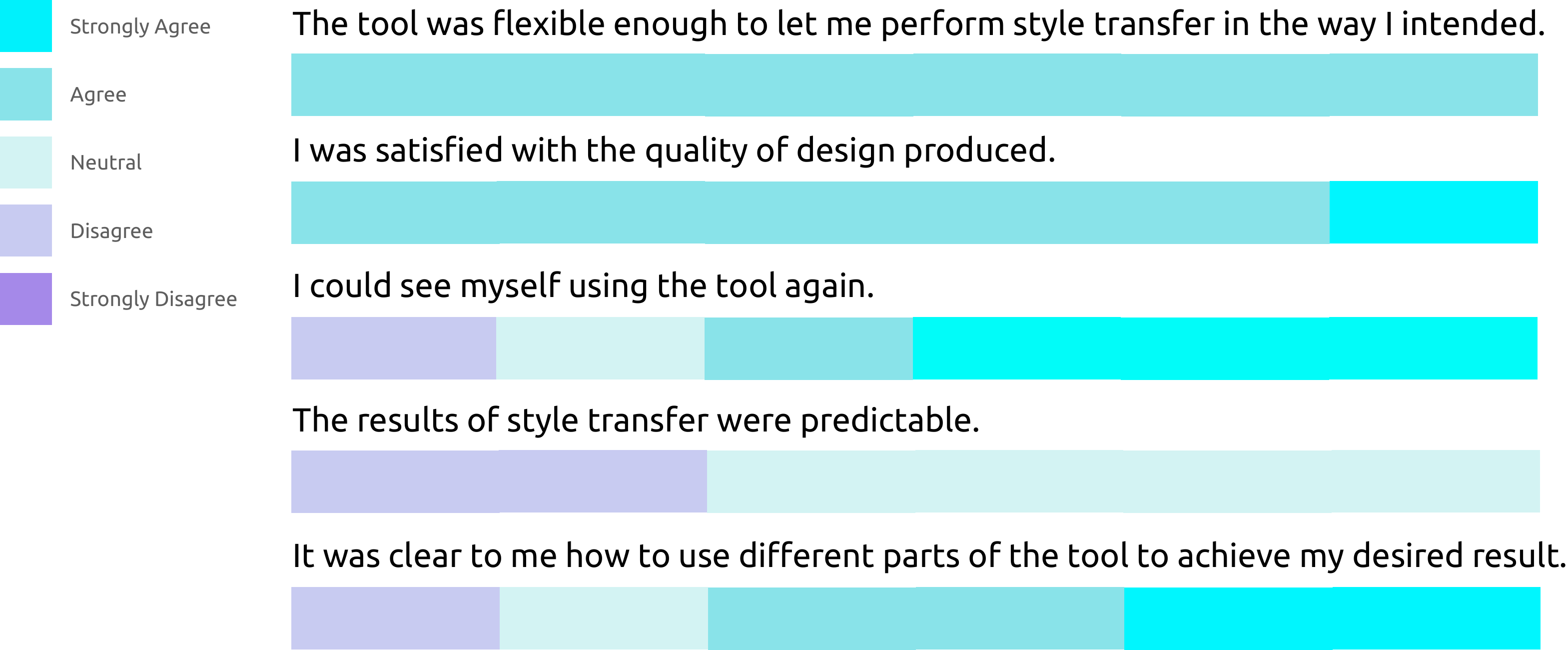}
   \Description[Enabling flexible style transfer (Likert responses)]{Study participants answered questions on flexibility, design satisfaction, using the tool again, predictability of the tool, and clarity around using the different parts of the tool to achieve the desired result.}      
  \caption{Summary of Likert survey data from designers \participant{D1-6}.}
  \label{fig:likert}
\end{figure}

\section{Discussion}

The success of VST demonstrates the value of two key design goals that are relevant as recommendations for other automation-powered correspondence-based transfer tools:
include the ability to flexibly \emph{tune generated design correspondences} (DG1)
and include the ability to flexibly \emph{customize what correspondences do} (DG2).

\textbf{Tuning Generated Design Correspondences.}
Providing powerful and convenient ways to tune correspondences avoids requiring users to make each mapping manually (DG1).
In VST, this functionality is represented by our \textit{Selecting Similar} feature, the ability to view and select elements sharing any of the same values in the customization pane, and the \textit{Similarity Threshold} feature (which lets users quickly preview selections).

\textbf{Customizing Correspondence Functions.}
Customizing a correspondence retains the flexibility of a manual approach, ensuring that designers still have control (DG2).
The domain will ultimately specify what is reasonable to transfer per correspondence.
Generally, the designer should be able to control what happens when two objects are linked.
In VST, we achieve this through our customization panel, where designers can copy, reset, and customize attribute values.
We also provide flexible ways to filter this list (e.g., by active selection and showing modified/all attributes).

\textbf{The Cost of Automation} --
One notable point in our results is that starting with the algorithm's output (\textit{Auto}) did not make replication easier.
In fact, the \participant{RD}s reported that starting with the automatically generated algorithm output was more difficult and less fun.
Simply throwing automation into existing tools and processes may backfire.
This is backed by our quantitative results: the \textit{Auto} designs, on average, required more work to style than the corresponding \textit{Basic} starting point
(\textit{Basic}: 265 operations, \textit{Auto}: 383 operations).
This is jarring, as applying the style transfer algorithm should have the opposite effect --- otherwise, why apply it at all?

First, applying a semi-correct transformation reduces cohesion in the design.
The lack of cohesion commonly found in \textit{Auto} designs reduces the efficiency of applying gestalt principles.
This makes selecting similar elements to style them together harder.
Second, the vast scope of the copied attributes may introduce new work.
Incorrectly changing an attribute does not create new work if it already needs to be changed.
However, if part of a \textsc{Source} style is not desired in the \textsc{Output} graphics, those attributes must be manually reset to their original \textsc{Target} value.
Current design software fails to support this type of style transfer interaction.
In contrast, VST features convenient ways to quickly select and explore element styles (double-clicking an element/selection, precision selection controls, visually selecting via the same attribute value).
Current correspondence algorithms do not seem to reduce the total work in style transfer otherwise.
This is especially true for more complex examples where correspondence accuracy is often lower.

\begin{table}[t]
\centering
    \caption{
    Replication work data -- 
    usage statistics averaged over replication tasks RT1-6 (see Fig.~\ref{fig:replication}).
    The \textit{Basic} and \textit{Auto} columns show aggregate data collected from the four expert replication designers (\participant{RD1-4}), while the \textit{VST} column shows data from the paper authors using VST to replicate designs.
    }
    \begin{tabular}{ll|rrr}
    &  & Basic  & Auto &  VST \\ 
    \hline
    Task Duration       & Mean & 532                 & 774                & \textbf{129}     \\
                        & S.D. & 341                 & 347                & 80               \\
    Work Operations     & Mean & 265.7               & 383.5              & \textbf{30.3}    \\
                        & S.D. & 167.8               & 159.2              & 18.9             \\
    Attribute Edits     & Mean & 80.0                & 113.1              & \textbf{13.0}    \\
                        & S.D. & 59.9                & 77.8               & 8.7              \\
    Selection Updates   & Mean & 185.7               & 270.4              & \textbf{17.3}    \\
                        & S.D. & 122.8               & 185.7              & 12.1             
    \end{tabular}
\label{table:replication}
\end{table}

\section{Limitations and Future Work}

\subsection{Limitations}

VST is not a general-use vector graphics editing platform.
The SVG standard is complex; even industry-standard platforms like Inkscape and Adobe Illustrator may render the same graphics differently.
Still, some missing features limited how useful VST was for designers in its current state.
Users wanted more advanced layering/z-reordering for sub-selections in complex design areas.
Additionally, the current correspondence structure usage limits elements to inheriting styles from one \textsc{Source} element unless manually mixed with other styles.

We also did not measure the impact of algorithm matching performance on this task.
Informally, study participants D1-6 updated the correspondence an average of six times per task, though our study instrumentation did not record the number of adjusted elements per update.
In Shin's prior work \cite{Shin2021MultilevelCV}, the average match accuracy was 95\
However, their evaluation \cite{Shin2021MultilevelCV} was performed with the \textsc{Source} as an element group within a \textsc{Target} design, rather than a separate design.
Explicitly varying the match quality and leveraging different matching techniques are opportunities for future work.
Another limitation of this work is the smaller scale of the surveyed designer population (10 unique designers across both studies).
For our design replication study, we worked with four expert designers.
While this smaller study size allowed us to deepen the level of feedback and data we gathered, future studies could evaluate a larger expert population to get additional feedback.
Future work could conduct a larger-scale study with more designers to potentially collect insights into a broader set of behaviors that designers exhibit.
Also, when comparing VST to other tools,
the authors have more awareness of the replication goal and task, which likely improves their relative performance.
Another evaluation could train experienced designers with VST and have them replicate graphics from the original study.

\subsection{Future Work}

Images can naturally add vibrancy to a design, though VST's style transfer only applies to vector graphics.
One future direction is sourcing vector styles directly from images.
\quoted{\participant{RD1:} It would be great to apply bitmap styling to my vector design. This use case is more common in my workflow.}
This requires converting the image to vector graphics or a novel style extraction technique.
Some features (e.g., colors) are simple to extract, while other features like paths, gradients, shapes, and fonts are potentially much harder to source from an image correctly.
For image-to-vector graphics conversion, some research methods \cite{Sbai2018VectorIG} and commercial tools \cite{adobeExpress} exist. However, these methods tend to optimize pixel-based similarity to the source image over a consistent output structure or element resolution.
The internal document complexity makes determining correspondences much more challenging.
Rasterizing vector graphics is a lossy process with no perfect inverse.
Still, given the ubiquity of image-based inspiration, a vector styling tool that uses images as a styling source is an exciting future direction.

Better correspondence algorithms may reduce the need for a corrective interface like VST.
Consider automatic speech transcription as an analogy: under a certain accuracy threshold, manually transcribing speech is easier than correcting a low-quality generated transcript.
The work required to fix the algorithm's output exceeds that of simply creating that same output manually.
There is room for improvement in design correspondence accuracy for vector graphics.
However, even with the best algorithm, some cases will still need manual tuning.
This ambiguity stems from the inherent subjectivity around \textit{good} style and varying designer tastes.

Primarily, our style transfer with this prototype addresses element size, font, stroke, and fill.
While designers can modify other features, this feature subset visually dominates the result.
A complete list of transferrable properties is in Appendix~\ref{svg-properties}.
Future work could serve as a larger-scale multi-design style linter or unification technique where many designs are edited simultaneously.
The design layout and structure are held constant throughout our style transfer process.
Applying the layout from source to target is an exciting and relevant next direction.

\section{Conclusion}

We presented a novel design tool called VST (Vector Style Transfer) for flexibly transferring styles across vector graphics designs.
We conducted two studies to investigate (1) how designers may use correspondence-based transfer tools like VST and (2) the potential of these tools in relation to traditional industry-standard design tools (e.g., Adobe Illustrator).
The first study, an open-ended style transfer evaluation, revealed that despite not previously using any similar tools, experienced designers could effectively transfer styles even across graphics independently created using other design tools.
\add{The second study, a preliminary design replication evaluation, suggests that tools like VST may reduce the time and work required to transfer styles across designs compared to traditional design tools.}
These expert designers also found directly editing automatically stylized graphics more difficult and tedious than the original baseline design templates.
This work provides two design recommendations for future design tools to support flexible user control: enable tuning generated design correspondences and customizing how these correspondences transform designs.

\bibliographystyle{ACM-Reference-Format}
\bibliography{./main.bib}
\appendix

\section{Methodology Details}
\label{sec:methodology}

\textbf{Style Transfer Evaluation} --
Four designers used Figma to generate their initial designs they brought into the study, while the other two designers used Adobe Illustrator.
After designers responded to the prompt, we hosted an hour-long Zoom session with each designer.
We instrumented the interface to log relevant events with timestamps (e.g., loading, saving, editing).
We sought to gather rich commentary and reflection from designers as they engaged with the prototype.
We invited designers to verbally share any thoughts on their experience and highlight any surprising interactions throughout the study.
While we recorded usage times per example, designers were not told this nor instructed to be as efficient as possible.
The designers moved on only after indicating satisfaction with the relative appearance of their stylized \textsc{Output} graphics.
Finally, designers answered a brief survey about their experience using VST, including Likert-scale (Fig.~\ref{fig:likert}) and open-ended questions.
Designers sent all styled designs and an interface usage log to the authors and received a \$30 Amazon gift card.

\textbf{Design Replication Evaluation} --
We conducted this study remotely over Zoom in a 3-hour session for each designer.
Unlike the style transfer evaluation, the \participant{RD}s were asked to work as swiftly and efficiently as possible.
Once the \participant{RD}s reported they were satisfied with the similarity between their replication graphics and the reference \textsc{Output} image, they would save their file and move on to the following example.
While the \participant{RD}s participated in our study, we recorded their screen, an audio log of the call, application edit history, and mouse activity.
From this data, we recorded the number of selections (including selection adjustments like shift-clicking or clicking the background to clear the selection) and attribute edits (per selection—so, for example, modifying the fill of a group counts as one edit).
We also recorded the time spent on each example task, measured from when all input files were opened to the last save of the output file.
Finally, the \participant{RD}s were shown, briefly used VST, and filled out a survey based on their experiences.
The \participant{RD}s received a variable Amazon gift card.
The amount was prorated based on their required completion time (rated at \$30/hour).

\section{Participant Backgrounds}
\label{sec:participants}

\textbf{Style Transfer Evaluation (D1-6)} -- 
We recruited designers via design-oriented email lists at a large research university.
Designers included undergraduates (4), Ph.D. students (1), and design professionals (1).
Each participating student had completed multiple design internships, bolstering their relevant experience.
Their preferred tools included Figma, Adobe Illustrator, and Canva.
They had an average of 4.7 years of design experience (2–10 years).

\textbf{Design Replication Evaluation (RD1-4)} --
We recruited from the same design community as before, now selecting only the most experienced designers.
All \participant{RD}s had professionally worked as designers. 
One was the instructor for a university course teaching students how to use Illustrator, and another held a residence in a design lab guiding student projects.
These designers had, on average, 6.5 years of design experience and used Illustrator daily.
None of these expert designers participated in the original study.

\add{

\section{Transferrable SVG Attributes}
\label{svg-properties}

The SVG attributes that VST can transfer are:
\textbf{(Color-Based)}
fill,
stroke,
strokeWidth,
textBackgroundColor,
\textbf{(Text-Based)}
lineHeight,
textAlign,
text (i.e., string content),
\textbf{(Font-Based)}
fontSize,
fontFamily,
fontStyle,
fontWeight,
and
\textbf{(General)}
opacity,
padding.
}

\section{Customization UI Techniques}
\label{interactions}

Also see our online demo: \url{https://berkeleyhci.github.io/vst/}

\begin{figure}[H]
        \centering
        
        \includegraphics[width=1\columnwidth]{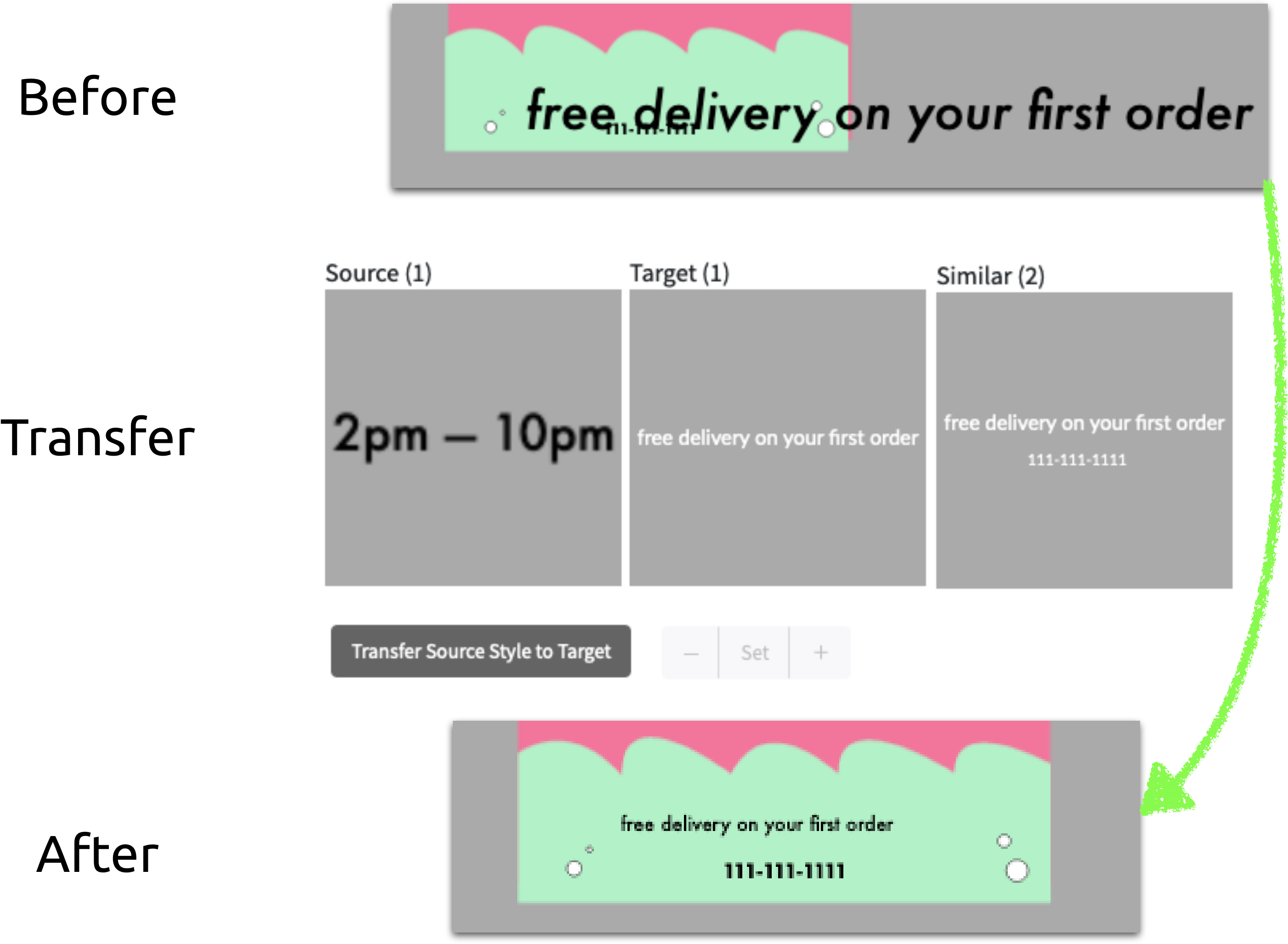}
      \Description[Updating the correspondence]{By specifying which source and targets to update, style can be redirected from the source to the target design.}  
        \caption{
The Customization UI shows the \textsc{Source} and \textsc{Target} selections and similar \textsc{Target} elements.
The similarity controls \texttt{[-/set/+]} can adjust the selection to the desired \textsc{Target} elements.
Once satisfied with the \textsc{Source-Target} mapping, pressing \textit{Transfer Source Style} will transfer all styles from the  \textsc{Source}  elements to the active  \textsc{Target}  selection.
      }
      \label{fig:system-match}
\end{figure}

\begin{figure}[H]
        \centering
        \includegraphics[width=1\columnwidth]{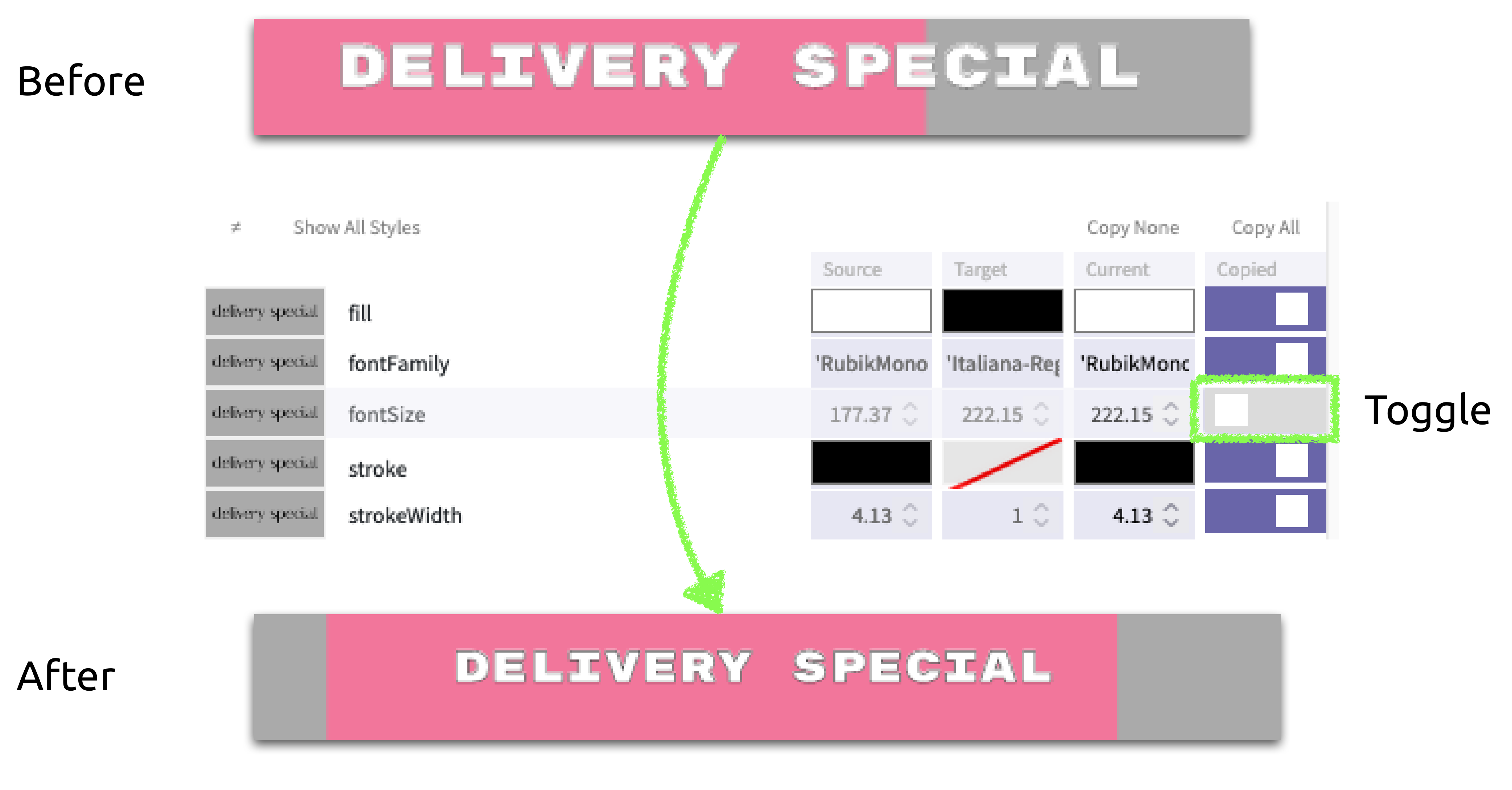}
        
 \Description[Updating styles copied from each correspondence]{Customization interface provides more fine-grained control over which styles are applied. Element style attributes can be copied, reset, or customized for each set of similar values.}       
        \caption{
        The Customization UI also provides fine-grained control over which styles to transfer.
        Element style attributes can be copied, reset, or customized for each set of similar values.
        This list can be filtered only to show styles for the current selection and only to show modified attributes.
        The UI also features the \textit{Copy All} and \textit{Copy None} buttons -- 
         \textit{Copy All} blindly copies all styles for every matched element (e.g., the fully automatic output), and \textit{Copy None} restores the \textsc{Output} graphics to the original \textsc{Target} state.
        }
        \label{fig:sys-style}
\end{figure}

\begin{figure*}[t]
        \centering
        \includegraphics[width=0.816\textwidth]{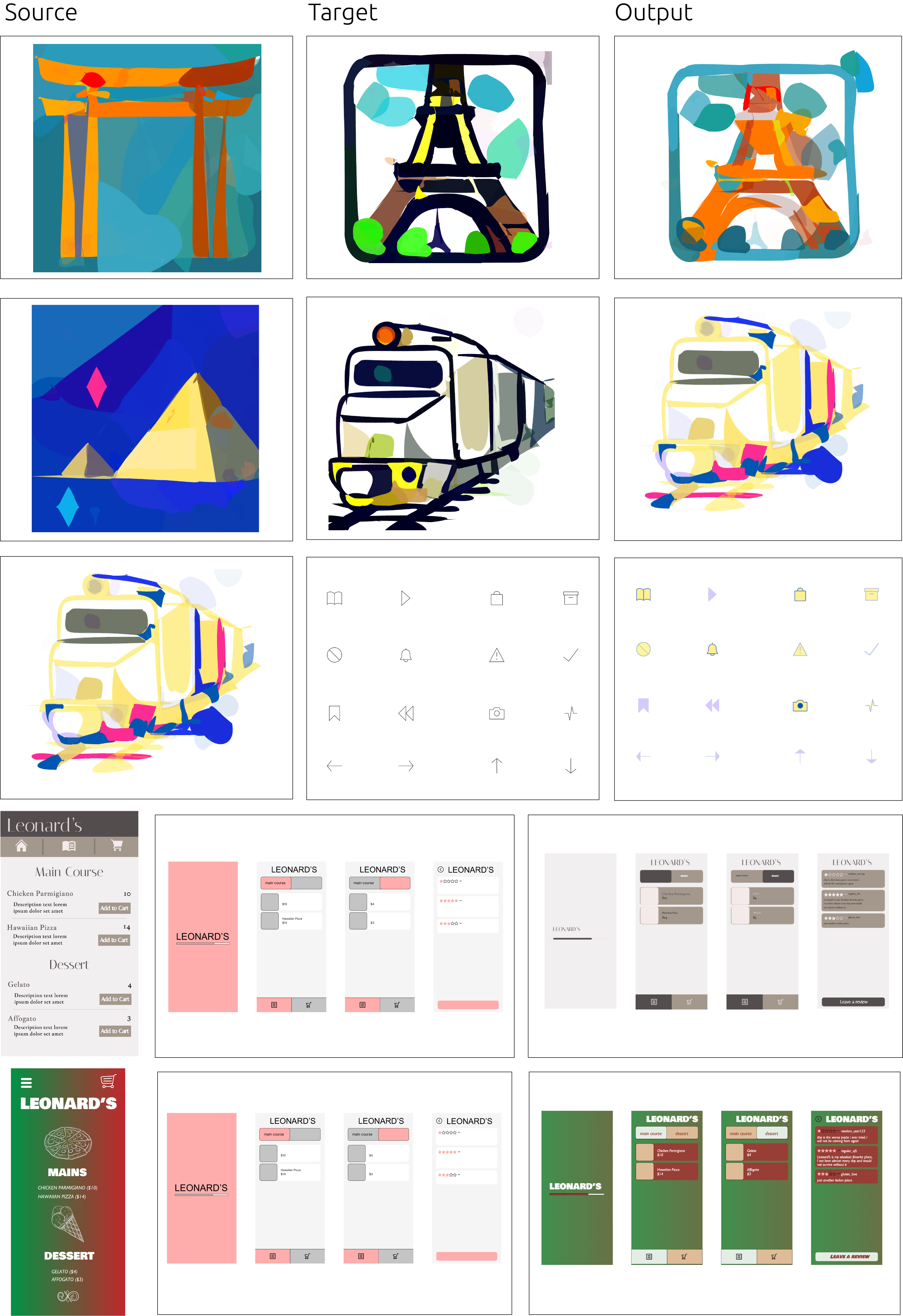}
        \Description[Additional graphics generated by transferring styles with VST.]{}
        \caption{Additional graphics generated by transferring styles with VST.}
        \label{fig:add-examples}
\end{figure*}

\end{document}